\pdfoutput=1
\documentclass{jfm}

\usepackage{subfigure}

\usepackage{graphicx}
\usepackage[framed,numbered,autolinebreaks,useliterate]{mcode}
\usepackage{natbib}
\usepackage{amsmath}
\usepackage{bm}
\usepackage{float} 
\usepackage{graphicx}
\usepackage{rotating}
\usepackage{latexsym}
\usepackage{tikz}
\usepackage{pgfplots}
\usepackage{algorithm}
\usepackage{algorithmic}

\pgfplotsset{
footnotesize/.style={
width=5cm,
height=,
legend style={font=\scriptsize},
tick label style={font=\scriptsize},
label style={font=\scriptsize},
title style={font=\scriptsize},
every axis title shift=0pt,
max space between ticks=15,
every mark/.append style={mark size=8},
major tick length=0.1cm,
minor tick length=0.066cm,
},
}

\pgfplotsset{
tiny/.style={
width=4cm,
height=,
legend style={font=\tiny},
tick label style={font=\tiny},
label style={font=\tiny},
title style={font=\footnotesize},
every axis title shift=0pt,
max space between ticks=12,
every mark/.append style={mark size=6},
major tick length=0.1cm,
minor tick length=0.066cm,
every legend image post/.append style={scale=0.8},
},
}

\pgfplotsset{a/.style={
        legend cell align=left,
        scaled ticks=true,
        width=6.5cm,
        height=6.5cm,
        xlabel={$a_1$},
        ylabel={$a_2$},
        xlabel near ticks,
        ylabel near ticks,
        ylabel style={rotate=-90},
    }
}

\pgfplotsset{K/.style={
        legend cell align=left,
        scaled ticks=true,
        width=6.5cm,
        height=6.5cm,
        xlabel={$t/T$},
        ylabel={$K$},
        xlabel near ticks,
        ylabel near ticks,
        ylabel style={rotate=-90},
    }
}

\pgfplotsset{State_space_n4/.style={
        State_space_n2,
        xlabel={$a_{i}$},
        ylabel={$a_{i+1}$},
        y label style={at={(-0.1,0.5)}}
    }
}

\pgfplotsset{Lines_black/.style={
        mark=none,
        line join=round,
        line width=0.5pt,
        color=black,
    }
}

\pgfplotsset{Lines_gray/.style={
        mark=none,
        line join=round,
        line width=2pt,
        color=gray,
    }
}

\pgfplotsset{Lines_red/.style={
        mark=none,
        line join=round,
        line width=0.5pt,
        color=red,
    }
}

\pgfplotsset{Lines_blue/.style={
        mark=none,
        line join=round,
        line width=0.5pt,
        color=blue,
    }
}

\ifCUPmtlplainloaded \else
  \checkfont{eurm10}
  \iffontfound
    \IfFileExists{upmath.sty}
      {\typeout{^^JFound AMS Euler Roman fonts on the system,
                   using the 'upmath' package.^^J}%
       \usepackage{upmath}}
      {\typeout{^^JFound AMS Euler Roman fonts on the system, but you
                   dont seem to have the}%
       \typeout{'upmath' package installed. JFM.cls can take advantage
                 of these fonts,^^Jif you use 'upmath' package.^^J}%
      }
  \else
  \fi
\fi

\ifCUPmtlplainloaded \else
  \checkfont{msam10}
  \iffontfound
    \IfFileExists{amssymb.sty}
      {\typeout{^^JFound AMS Symbol fonts on the system, using the
                'amssymb' package.^^J}%
       \usepackage{amssymb}%
         \let\leq=\leqslant
         \let\geq=\geqslant
      }{}
  \fi
\fi

\ifCUPmtlplainloaded \else
  \IfFileExists{amsbsy.sty}
    {\typeout{^^JFound the 'amsbsy' package on the system, using it.^^J}%
     \usepackage{amsbsy}}
    {\providecommand\boldsymbol[1]{\mbox{\boldmath $##1$}}}
\fi

\newsavebox{\astrutbox}
\sbox{\astrutbox}{\rule[-5pt]{0pt}{20pt}}

\title[Lyapunov stable Galerkin models of post-transient incompressible flows]{Lyapunov stable Galerkin models of post-transient incompressible flows}

\author[M.J.\ Balajewicz]%
{M\ls A\ls C\ls I\ls E\ls J\ns J.\ns B\ls A\ls L\ls A\ls J\ls E\ls W\ls I\ls C\ls Z$^1$%
  \thanks{Email address for correspondence: maciej.balajewicz@stanford.edu}}

\affiliation{$^1$Department of Aeronautics and Astronautics, Stanford University, Stanford, CA 94305, USA}

\pubyear{2010}
\volume{650}
\pagerange{119--126}
\begin{document}

\maketitle

\begin{abstract}
A method for deriving provably stable low-dimensional Galerkin models of post-transient incompressible flows is introduced. The proposed approach involves an iterative procedure for expansion modes that satisfy Lyapunov stability in the neighborhood of a fixed point. The approach is demonstrated using two prototypical flow configurations: a two-dimensional mixing layer, and two-dimensional flow inside a lid-driven cavity. This new methodology can be a building block in an effort to develop more accurate and more robust low-dimensional models of incompressible flows.
\end{abstract}


\section{Introduction}
The Proper Orthogonal Decomposition (POD) and Galerkin projection form a popular model order reduction approach for incompressible flows~\citep{Noack:2011,Cordier:2013}. Despite many recent advances, low-dimensional POD-Galerkin models tend to be unstable and empirical modifications are required to achieve long-term boundedness~\citep{Bailon:2012,Iliescu:2012,Noack:2012b,Wang2012cmame}. In this paper, a novel approach that addresses this issue is presented. Specifically, the proposed approach yields Galerkin models with Lyapunov stable fixed points. Trajectories in the neighborhood of these fixed points are, therefore, guaranteed to remain bounded for all time. Moreover, these trajectories are demonstrated to reproduce surprisingly well the dynamics of the Navier-Stokes attractor.\\
\indent
This paper is organized as follows. In \S 2 a Lyapunov candidate function is defined and a method for deriving Galerkin models that satisfy Lyapunov stability is summarized. In \S 3 the proposed approach is demonstrated on two prototypical flow configurations. Finally, in \S 4, the main results are summarized and future prospects laid out.
\section{Methodology}
\subsection{Lyapunov candidate function}
\label{sec:Methodology}
In this section a Lyapunov candidate function for Galerkin models of post-transient incompressible flows is derived. A Galerkin model approximates the flow velocity $\bm{u}(\bm{x},t)$ with a finite dimensional subspace~\citep{Holmes:2012}:
\begin{equation}
\label{Eqn:GE}
    \bm{u}(\bm{x},t) := \bm{u}_0(\bm{x}) + \sum\limits_{i=1}^n {a_i (t) \> \bm{u}_i (\bm{x})}
\end{equation}
where $\bm{u}_0$ is the temporal mean flow and the expansion modes $\bm{u}_i$ arise from a proper orthogonal decomposition (POD) of solution snapshots. A Galerkin projection yields a set of evolution equations for the modal amplitudes:
\begin{equation}
    \label{Eqn:GS}
    \frac{d a_i}{{dt}}  = C_i + \sum\limits_{j = 1}^n L_{ij} a_j + \sum\limits_{j,k = 1}^n Q_{ijk} a_j a_k.
\end{equation}
By the transformation $a_i = z_i + h_i$, Eq.~\eqref{Eqn:GS} can be shifted to the origin yielding the system:
\begin{equation}
    \label{Eqn:GS_z}
    \frac{d z_i}{{dt}}  = \sum\limits_{j = 1}^n A_{ij} z_j + \sum\limits_{j,k = 1}^n Q_{ijk} z_j z_k
\end{equation}
with fixed point $\bm{z}=\bm{0}$. The matrix $\mathsfbi{A}$ is the ``centered'' linear Galerkin tensor:
\begin{equation}
    \label{Eqn:A}
    A_{ij} : = L_{ij}  + \sum\limits_{k = 1}^n {(Q_{ijk}  + Q_{ikj} )h_k }.
\end{equation}
Let $V$ be a Lyapunov candidate function:
\begin{equation}
    \label{Eqn:En_POD}
    V(\bm{z}) := \frac{1}{2} \sum\limits_{i=1}^n z_i^2
\end{equation}
that is locally positive definite, i.e. $V(\bm{0})=0$, and $V(\bm{z}) > 0 \, \forall \bm{z}  \in U \backslash \{ \bm{0}\}$ with $U$ being a neighborhood region around $\bm{z}=0$. This Lyapunov function is related to the turbulent kinetic energy (TKE) of the flow $E:={1 \mathord{\left/
 {\vphantom {1 2}} \right.
 \kern-\nulldelimiterspace} 2}\sum\nolimits_{i = 1}^n {a_i^2 } $. The time derivative of V is:
\begin{equation}
    \frac{dV} {dt} = \nabla V \cdot \frac{\partial z} {\partial t}
    = \sum\limits_{j,k = 1}^n A_{ij}  z_j z_k.
\end{equation}
For a large class of boundary conditions the quadratic term can be shown to be energy-preserving (i.e. $Q_{ijk} + Q_{ikj} + Q_{jik} + Q_{jki}  + Q_{kij} + Q_{kji} = 0$) and therefore, does not contribute to the time derivative of $V$~\citep{Kraichnan:1989,Schlegel:2013}. \\
\indent Finally, by Lyapunov's direct method for stability, the fixed point $\bm{z}=\bm{0}$ (equivalently $\bm{a}=\bm{h}$) is:
 \begin{enumerate}
   \item \textit{asymptotically stable} if $\mathsfbi{A}$ is negative definite~\citep{Goulart:2012,Schmid:2001,Schlegel:2013}. This is true if and only if the Hermitian part $\mathsfbi{A}_{H}:=(\mathsfbi{A} + \mathsfbi{A}^{ \rm T})/2$ is negative definite. Or, equivalently, if $ {\eta (\mathsfbi{A}_H)} < 0$ where $\eta (\mathsfbi{A}_H) := \max \{ \lambda _i \} $ is the spectral abscissa of the matrix $\mathsfbi{A}_H$ and $\lambda _i$ are its eigenvalues.
   \item \textit{Lyapunov stable} if the Hermitian part $\mathsfbi{A}_{H}:=(\mathsfbi{A} + \mathsfbi{A}^{ \rm T})/2$ vanishes, i.e. $\mathsfbi{A}_{H}~=~\mathsfbi{0}_{n \times n}$.
 \end{enumerate}

In this work, primary interest is high-Reynolds-number flows and thus, the only physically representative stability characteristic is Lyapunov stability. The basis idea of the proposed approach is to search for a set of fluid modes that provide a Lyapunov stable Galerkin dynamical system. This can be accomplished via the generalized Galerkin approximation first introduced in~\citet{Balajewicz_JFM_2013} and summarized in the following section.
\subsection{Generalized Galerkin approximation}
\label{sec:Problem definition}
In the generalized Galerkin model approximation, the velocity field of the flow is approximated by expansion modes $\left\{ \bm{u}_i (\bm{x}) \right\}_{i=1}^{n}$ that are linear superpositions of $N$ ($N>n$) POD modes. Thus, one can write:
\begin{equation}
\label{eqn:generalized_Galerkin_approximation}
    \bm{u}_i(\bm{x}) := \sum\limits_{j = 1}^{N} {X_{ji} \bm{u}_j^* (\bm{x})}
\end{equation}
where $\left\{ \bm{u}^*_i (\bm{x}) \right\}_{i=1}^{N}$ are the POD modes and $\mathsfbi{X} \in \mathbb{R}^{N \times n}$ is a orthonormal ($\mathsfbi{X}^{\rm T} \mathsfbi{X} = \mathsfbi{I}_{n \times n}$) transformation matrix. The Galerkin system tensors can be all expressed as a function of $\mathsfbi{X}$ as follows
\begin{subequations}
\label{eqn:new_Galerkin_tensors}
    \begin{align}
        Q_{ijk} &=
            \sum\limits_{p,q,r = 1}^N {X_{pi} Q_{pqr}^* X_{qj}  X_{rk} } \hspace{5mm} i,j,k=1,\cdots,n,  \\
            \mathsfbi{L} &= \mathsfbi{X}^{\rm T} \mathsfbi{L}^* \mathsfbi{X}, \\
            \mathsfbi{C} &= \mathsfbi{X}^{\rm T} \mathsfbi{C}^*
    \end{align}
\end{subequations}
where
$\mathsfbi{C}^* \in \mathbb{R}^{N}$,
$\mathsfbi{L}^* \in \mathbb{R}^{N \times N}$ and
$\mathsfbi{Q}^* \in \mathbb{R}^{N \times N \times N}$,
are the Galerkin system coefficients corresponding to POD modes $\left\{ \bm{u}^*_i (\bm{x}) \right\}_{i=1}^{N} $. The new modes are expected to capture a lower proportion of the turbulent kinetic energy compared to the optimal POD modes. This loss of optimality is quantified by the optimality ratio $\xi  := {{\sum\nolimits_{i = 1}^n {(\sigma _i )^2 } } \mathord{\left/
 {\vphantom {{\sum\nolimits_{i = 1}^n {(\sigma _i )^2 } } {\sum\nolimits_{i = 1}^n {(\sigma _i^ *  )^2 } }}} \right.
 \kern-\nulldelimiterspace} {\sum\nolimits_{i = 1}^n {(\sigma _i^ *  )^2 } }}$ where $\sigma*$ are the POD eigenvalues and $\sigma_i : = (\mathsfbi{X}^{\rm T} diag(\sigma ^*)\mathsfbi{X})_{ii}$ are the eigenvalues associated with the new modes. The optimality ratio is bounded $0 \leqslant \xi  \leqslant 1 $ for all orthonormal $\mathsfbi{X}$ since $\sigma _1^ *   \geqslant \sigma _2^ *   \geqslant  \ldots  \geqslant \sigma _N^ *$.
\subsection{Construction of Lyapunov stable Galerkin models}
\label{sec:Problem definition}
In this section, a method for deriving Lyapunov stable Galerkin models using the generalized Galerkin approximation is developed. The main idea is to search for modes (i.e. find $\mathsfbi{X}$ ) that yield a Lyapunov stable Galerkin dynamical system. A procedure for numerically identifying these solutions is summarized as follows.\\
\indent Let $f(\mathsfbi{X})$ be an operator of $\mathbb{R}^{N \times n}$ into $\mathbb{R}^{n \times n}$ whose output is the Hermitian part of $\mathsfbi{A}$. Solutions of $f(\mathsfbi{X})=0$ are found using the generalized Newton's method:
\begin{equation}
\label{eqn:Newton}
    \mathsfbi{X}^{k + 1}  = \mathsfbi{X}^k  - f'(\mathsfbi{X}^k )^ +  f(\mathsfbi{X}^k ) \hspace{5mm} k=0,\cdots,K
\end{equation}
where  $\mathsfbi{X}^0  = \left[ {\begin{array}{*{20}c}
   {\mathsfbi{I}_n } & {\mathsfbi{0}_{N - n} }  \\
 \end{array} } \right]^{\rm T}
$ and $f'(\mathsfbi{X}^k )^ +$ denotes the Moore-Penrose pseudoinverse of the Jacobian. The operator $f$ involves several steps. First, the transformation is constrained to be orthonormal using the matrix square root $\mathsfbi{X} = \mathsfbi{X}(\mathsfbi{X}^{\rm T} \mathsfbi{X})^{ - 1/2}$. Second, the new Galerkin tensors $\mathsfbi{C}$, $\mathsfbi{L}$, and $\mathsfbi{Q}$ are computed using Eq.~\eqref{eqn:new_Galerkin_tensors}. Third, a fixed point of the new Galerkin system is found using Newton's method with initial guess $\bm{h}=\bm{0}$. Fourth, the system is centered around this fixed point via the transformation $z_i = a_i + h_i$ yielding the centered Galerkin tensor $\mathsfbi{A}$. Finally, the Hermitian part $\mathsfbi{A}_H:=(\mathsfbi{A}+\mathsfbi{A}^{\rm T})/2$ is returned. Equation~\eqref{eqn:Newton} is solved until $\| \mathsfbi{A}_H \|_F < {\rm Tol}$.
\subsection{Numerical implementation}
A numerical implementation of the proposed approach in MATLAB is provided in the Appendix. The function \verb=fsolve= is used to solve Equation~\eqref{eqn:Newton} using the Levenberg-–Marquardt algorithm. Several implementation details are worthy of mention. First, although the Jacobian can approximated using finite-differences, this requires $N\times n$ evaluations of $f(\mathsfbi{X}^k )$. For large modes these evaluations are computationally prohibitive. Instead, the Jacobian is approximated using an surrogate function $\tilde{f}(\mathsfbi{X}^k )$. The surrogate function outputs the Hermitian part of the \textit{uncentered} Galerkin tensor $\mathsfbi{L}$. This approximation significantly decreases computational costs and, for the specific flow configurations tested here, does not significantly effect convergence properties. Secondly, the method introduced in this paper does not guaranteed existence of solutions, i.e. $f(\mathsfbi{X})=0$. However, numerical experiments summarized in this paper suggest that a rule of thumb of $N \approx 10n$ yields solutions and good overall performance. Finally, the proposed algorithm assumes that the provided quadratic Galerkin tensor $\mathsfbi{Q}^*$ is energy preserving, i.e. $Q_{ijk}^* + Q_{ikj}^* + Q_{jik}^* + Q_{jki}^*  + Q_{kij}^* + Q_{kji}^* = 0$
\section{Applications}
In this section the proposed approach is demonstrated on two prototypical flow configuration: a two-dimensional mixing layer, and two-dimensional flow inside a lid-driven cavity.
\subsection{Two-dimensional mixing layer}
The data base corresponds to a direct numerical simulation (DNS) of an isothermal two-dimensional mixing layer. The numerical algorithm is the same as that employed previously for studies on jet noise sources~\citep{Cavalieri:2011}. The inflow mean streamwise velocity profile is given by a hyperbolic tangent profile
    \begin{equation}
        \overline{u}(y)=U_2+ \Delta U  \> \left[ \frac{1 + \tanh(2y)}{2} \right],
    \end{equation}
with $\Delta U=U_1-U_2$ the velocity difference across the mixing layer, where $U_1$ and $U_2$ are the initial velocity above and below, respectively. The flow Reynolds number is $Re=\delta_{\omega} \Delta U/ \nu_{a}=500$ where the subscript $\left(\cdot\right)_a$ indicates a constant ambient quantity and $\delta_{\omega}$ is the inflow boundary layer thickness. The simulation is first initialized over $330\,000$ time steps ($\Delta t=0.002$) the data base is then generated: $1\,093\,695$ iterations corresponding to $2000$ snapshots. A total of $100$ POD modes are derived from this database. 
\subsection{Two-dimensional lid-driven cavity}
The incompressible, two-dimensional flow inside a square cavity driven by a prescribed lid velocity, $\bm{u}_{lid}=(1-x^2)^2$ is considered. The flow Reynolds number is $3 \times 10^4$ defined with respect to the maximum velocity of the lid and the width of the cavity. The Navier-Stokes equations are discretized in space using Chebyshev polynomials. The convective nonlinearities are handled pseudo-spectrally and the Chebyshev coefficients are derived using the Fast Fourier Transform (FFT). This system is integrated in time using a semi-implicit, second-order scheme. The simulation is first initialized over $100\,000$ time steps ($\Delta t=1 \times 10^{-3}$) and then the data base is then generated: $25\,000$ iterations corresponding to $2500$ snapshots. A total of $100$ POD modes are derived from this database.
\subsection{Low-dimensional Galerkin models}
\begin{figure}
\centering
    \includegraphics{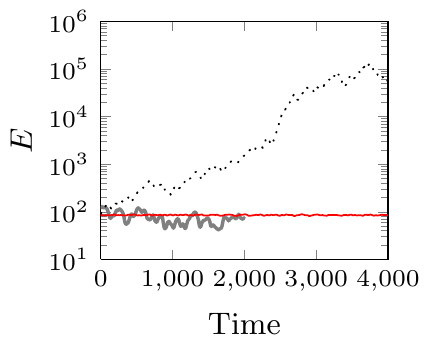}
    \includegraphics{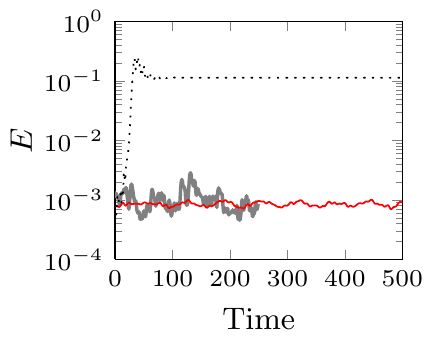}
\caption{Evolution of the instantaneous turbulent kinetic energy of the mixing layer (left), and lid-driven cavity (right) as predicted by a standard POD Galerkin model (dotted lines), a Lyapunov stable Galerkin model (red lines), and the DNS (grey lines).}\label{fig:E}
\end{figure}

\begin{figure}
\centering
    \includegraphics{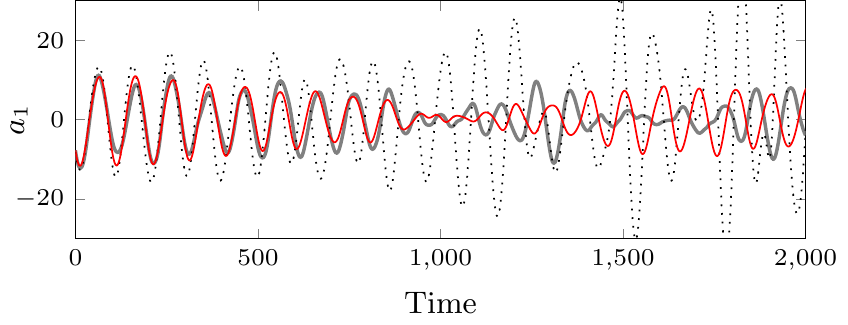}
    \includegraphics{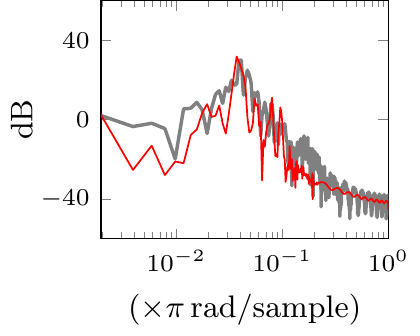}
    \includegraphics{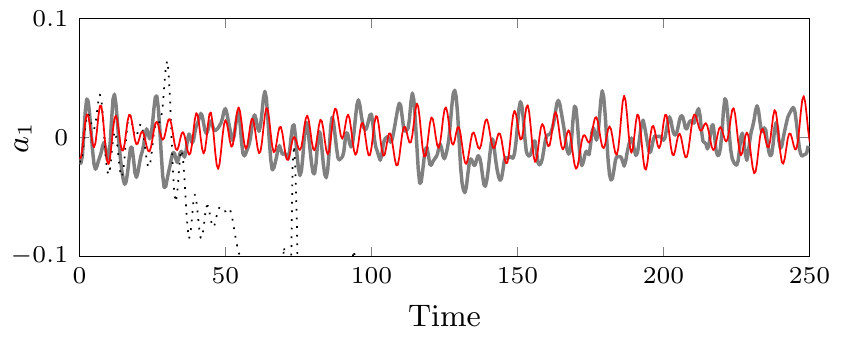}
    \includegraphics{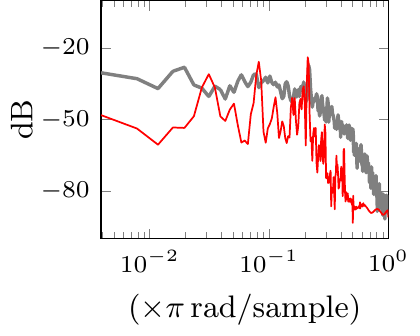}
\caption{Evolution of the first modal coefficient $a_1(t)$ and its power spectral density (PSD) as predicted by  standard POD Galerkin models (dotted lines), Lyapunov stable Galerkin models (red lines), and the DNS (grey lines). Mixing layer (top), lid-driven cavity (bottom).}\label{fig:a1}
\end{figure}

Both flow configuration are approximated using two, $n=10$ Galerkin models. The first model corresponds to a standard POD-Galerkin model using the first $10$ most energetic POD modes. The second model corresponds to a $n=10$ Lyapunov stable Galerkin model derived via a linear superposition of $N=100$ POD modes. The optimality ratio $\xi$ between the new modes and the POD modes is $0.8882$ for the mixing layer and $0.9014$ for the lid-driven cavity. Evolution of the turbulent kinetic energy predicted by these models is illustrated in Figure~\ref{fig:E}. As expected, the derived Lyapunov stable models remain strictly bounded while the standard POD-Galerkin models significantly overpredict the energy. In Figure~\ref{fig:a1}, evolution of the first modal coefficient $a_1(t)$ and its power spectral density are illustrated. For both flow configurations, trajectories of the new Galerkin models are surprisingly representative of the DNS simulation.


%
%
%
%

\section{Conclusions and prospects for future work}
In the proposed approach, Lyapunov stable Galerkin models are found using modes that are linear superpositions of POD modes. Specifically, the new modes are found such that the centered Galerkin tensor vanishes. The resulting modes are guaranteed to remain bounded for all time in the neighbourhood of the Lyapunov stable fixed point. The results presented in this paper are surprising for several reasons. First, solutions are demonstrated to exist for a wide range of Reynolds number; $500$ for the mixing layer and $3 \times 10^4$ for the driven cavity. Second, the trajectories of the derived models accurately approximate the dynamics of the the Navier-Stokes attractor. This is especially surprising given the small size of the models. Several opportunities for future work are summarized here. \\
\indent The derived Lyapunov stable Galerkin models constitute energy-conservative approximations of the intrinsically dissipative Navier-Stokes attractor. This property can be viewed as both a weakness and a strength of the proposed approach. It is a weakness because the derived models can not be expected to perform well off the attractor. The models simply inherit the energy of the initial condition for all time. On the other hand, this property is a strength because it opens the possibility of augmenting the conservative models with a empirical dissipative term. For example, the centered Galerkin model can be augmented as follows: 
\begin{equation}
    \label{Eqn:GS_z}
    \frac{{dz_i }}
    {{dt}} = \sum\limits_{j = 1}^n {A_{ij} } z_j  + \sum\limits_{j,k = 1}^n {Q_{ijk} } z_j z_k  + \alpha_i (V^* - V({\bm{z}}))z_i
\end{equation}
where $V^*$ is the energy on the attractor and $V({\bm{z}})$ is the instantaneous energy. The free parameters $\alpha_i$ can be tuned empirically to match the desired overall dissipation. Most importantly, once the trajectories return to the attractor the extra empirical term vanishes and the unmodified Galerkin model is recovered.\\
\indent The proposed approach operates on the level of the Galerkin dynamical system, i.e. the quadratic ODEs derived via a Galerkin projection of the Navier-Stokes equation. Therefore, since only the Galerkin tensors are required, the approach is independent of the algorithm used to derive the fluid modes. POD modes were used in this work but other algorithms such as dynamic mode decomposition (DMD) could be utilized.\\
\indent Another opportunity for future work involves weakening the condition of vanishing Hermitian. Instead of searching for a conservative model, a non-conservative model would be identified that best fits the data. This would involve a search for $\mathsfbi{X}$ such that the distribution of the eigenvalues of $\mathsfbi{A}_H$ yields an accurate and dissipative Galerkin model of the Navier-Stokes attractor. Since POD modes are biased toward the energy containing scales of the flow, the approach would involve decreasing the magnitude of positive eigenvalues of $\mathsfbi{A}_H$ and increasing the magnitude of negative eigenvalues. More generally, the method could be modified to search for monotonically attracting trapping regions introduced by~\citet{Schlegel:2013}.  \\
\indent A final opportunity for future work involves modifying the proposed approach to a constrained minimization problem. In addition to providing Lyapunov stable models, the goal would be to find modes that are as close as possible to the optimal POD modes:
 \begin{equation}
    \begin{aligned}
    & \underset{\mathsfbi{X} \in \mathbb{R}^{N \times n}}{\text{minimize}}
    & & \displaystyle 1 - \xi \\
    & \text{subject to}
    & &  \| \mathsfbi{A}_H \|_F = 0.
    \end{aligned}
\end{equation}
\section*{Acknowledgements}
    \label{sec:Acknowledgements}
The author is particularly grateful to Bernd Noack from the Institute PPRIME, and Guillame Daviller from CERFACS for making available the mixing layer data set.
\clearpage
\appendix
\section{MATLAB implementation}\label{appA}
The following is a simple MATLAB implementation of the proposed approach. For the sake of clarity, several global variables are utilized and these must be defined and present in the workspace. Specifically, the POD Galerkin tensors are required in the following format: \verb=C_star(i)=~$=C_{i}^*$, \verb=L_star(i,j)=~$=L_{ij}^*$ and \verb=Q_star{i}(j,k)=~$=Q_{ijk}^*$ for $i,j,k=1,\cdots,N$. The output of the function \verb=Lyap_stab_Galerkin= is the orthonormal transformation matrix $\mathsfbi{X} \in \mathbb{R}^{N \times n}$ defining the expansion modes that generate a Lyapunov stable Galerkin model.
\begin{lstlisting}
function [X] = Lyap_stab_Galerkin(n)
global n N C_star L_star Q_star
    x0 = eye(n,n);
    x0(N,n) = 0;
    options = optimset('algorithm','levenberg-marquardt',...
    'Jacobian','on','TolX',1e-9,'TolFun',1e-9,...
    'MaxFunEvals',1000,'MaxIter',100,'Display','iter',...
    'ScaleProblem','Jacobian');
    [x,res,EXITFLAG,OUTPUT] = fsolve({@f,@Jacobian},x0,options);
    X = x*(x'*x)^(-1/2);
end
\end{lstlisting}

\begin{lstlisting}
function [A_H] = f(x)
global n
    X = x*(x'*x)^(-1/2);
    [C,L,Q,a] = make_tensors(X);
    fun = @(a)Galerkin_Navier_Stokes(a,n,C,L,Q);
    options = optimset('TolFun',1e-12,'Diagnostics','off',...
    'Display','off');
    [h,~,~,~] = fsolve(fun,zeros(n,1),options);
    A = zeros(n,n);
    for i=1:n; A(i,:) = L(i,:) + h'*(Q{i}+Q{i}'); end
    A_H = (A + A')/2; A_H = A_H(:);
end
\end{lstlisting}

\begin{lstlisting}
function [da] = Galerkin_Navier_Stokes(a,n,C,L,Q)
da = zeros(n,1);
    for i=1:n; da(i) = C(i,1) + L(i,:)*a + a'*Q{i}*a; end
end
\end{lstlisting}

\begin{lstlisting}
function [J] = Jacobian(x)
global n N
    R = f_tilde(x); J = zeros(n*n,n*N);
    for i=1:n*N
        e = 1e-10; d_x = zeros(n*N,1); d_x(i) = e;
        J(:,i) = (f_tilde(x + reshape(d_x,N,n))-R)/e;
    end
end
\end{lstlisting}
\newpage
\begin{lstlisting}
function [L_H] = f_tilde(x)
global L_star
    X = x*(x'*x)^(-1/2);
    L = X'*L_star*X;
    L_H = (L + L')/2; L_H = L_H(:);
end
\end{lstlisting}

\begin{lstlisting}
function [C,L,Q,a] = make_tensors(X)
global n N C_star L_star Q_star
    C = X'*C_star; L = X'*L_star*X; Q = cell(n,1);
    for  i = 1:n
        for  j = 1:n
            for  k = 1:n
                Q{i}(j,k) = X(1,i)*(X(:,j)'*Q_star{1}*X(:,k));
                for p = 2:N;
                    Q{i}(j,k) = Q{i}(j,k) + X(p,i)*(X(:,j)'*Q_star{p}*X(:,k));
                end
            end
        end
    end
end
\end{lstlisting}

\bibliographystyle{jfm}

\end{document}